\documentclass[iicol, sn-basic, Numbered]{sn-jnl}

\usepackage[T1]{fontenc}
\usepackage[utf8]{inputenc}
\usepackage{lmodern}    
\usepackage{amsmath,amssymb,amsfonts}
\usepackage{bm}     
\usepackage{physics}
\usepackage{graphicx} 
\usepackage{microtype} 
\usepackage{tabularx}
\usepackage{booktabs}
\usepackage{hyperref}
\usepackage{natbib}

\newcommand{\ub}[0]{^{(B)}}
\newcommand{\uq}[0]{^{(Q)}}
\newcommand{\wc}[0]{w_{\text{couple}}}

\begin{document}

\title{Quantum error correction and biological error correction:\\ A structural analogy between qubits and neurons}

\author*[1,2]{\fnm{Ian} \sur{Whitehouse}}\email{ianjw@umd.edu}
\author*[1]{\fnm{An\i l} \sur{Zengino\u{g}lu}}\email{anil@umd.edu}
\author*[3]{\fnm{Franz} \sur{Klein}}\email{fklein@umd.edu}
\author[2]{\fnm{Mohe} \sur{Edeen Abu Maizer}}
\author[2]{\fnm{Wilson} \sur{Smith}} % any umd math/cs affil works
\author[5]{\fnm{Skylar} \sur{Chan}}
\author[4]{\fnm{Siri} \sur{Duddella}}
\author*[1, 4]{\fnm{Wolfgang} \sur{Losert}}\email{wlosert@umd.edu}

\affil[1]{\orgdiv{Institute for Physical Science and Technology}, \orgname{University of Maryland}, \orgaddress{\city{College Park}, \postcode{20742}, \state{MD}, \country{USA}}}
\affil[2]{\orgdiv{Department of Computer Science}, \orgname{University of Maryland}, \orgaddress{\city{College Park}, \postcode{20742}, \state{MD}, \country{USA}}}
\affil[3]{\orgdiv{National Quantum Laboratory}, \orgname{University of Maryland}, \orgaddress{\city{College Park}, \postcode{20742}, \state{MD}, \country{USA}}}
\affil[4]{\orgdiv{Department of Physics}, \orgname{University of Maryland}, \orgaddress{\city{College Park}, \postcode{20742}, \state{MD}, \country{USA}}}
\affil[5]{\orgdiv{School of Medicine}, \orgname{University of Maryland}, \orgaddress{\city{Baltimore}, \postcode{21201}, \state{MD}, \country{USA}}}
%\affil[6]{\orgdiv{Laboratory of Cardiovascular Sciences}, \orgdiv{National Institute on Aging}, \orgaddress{\city{Baltimore}, \postcode{20892}, \state{MD}, \country{USA}}}

% \begin{abstract}
\abstract{We draw a structural analogy between quantum error correction (QEC) and error handling in neural circuits, with respect to their redundant encodings and constraint-based inferences. In QEC, logical information is embedded in a protected codespace within a larger Hilbert space. A set of commuting checks (e.g. stabilizer constraints) is repeatedly evaluated to produce an error syndrome that identifies which constraints were violated without directly revealing the logical state. A decoder then maps the syndrome to a recovery operation that returns the system to the codespace and suppresses logical failure below a threshold.

Neural circuits exhibit error-control strategies that can be viewed through a related biological error correction (BEC) pattern: information is distributed across multiple neurons (redundant encoding), yielding reliable collective activity from error-prone unit operations of individual neurons. The structural analogy with QEC raises the question whether collective activity may be constrained on lower-dimensional manifolds (a biological codespace), allowing recurrent circuit dynamics and mismatch signals to function as syndrome-like indicators of constraint violations, driving fast corrective dynamics and slower adaptive updates.

Our structural analogy also suggests that new insights into brain-inspired algorithms for collective information processing may inform novel QEC approaches. We perform numerical experiments using simplified models of qubit and neural population dynamics to illustrate the analogy.
}
% \end{abstract}


\maketitle

\section{Introduction}

How can we perform reliable computation with unreliable parts? This fundamental question has shaped both the theory of computing and our understanding of neural systems for decades. Von Neumann's classic lectures on ``Probabilistic Logics and the Synthesis of Reliable Organisms from Unreliable Components'' formulated this question for noisy logical elements and for neural networks, showing that redundancy and majority voting can ensure reliable computation, provided the component error rate lies below a threshold~\cite{vonNeumann1956}. This insight shapes both classical and quantum error-correcting codes today.

Over the last three decades, quantum information has provided new perspectives into error correction. Shor's discovery of a QEC code that protects a logical qubit from arbitrary single physical qubit errors~\cite{Shor1995} and the later development of the stabilizer formalism and topological codes~\cite{NielsenChuang2010, Terhal2015}, established that reliable quantum computation is possible for noisy physical qubits within a specific error threshold. In these constructions, logical information is encoded nonlocally in a subspace of a larger Hilbert space called the \emph{codespace}, and ancillary degrees of freedom help extract error syndromes without collapsing the encoded logical state. QEC thus implements von Neumann's vision in the quantum domain by using structured redundancy and constrained dynamics to protect fragile information. This separation between protected information (the logical degrees of freedom) and measured information (the syndrome) is the core principle of QEC.

In this paper, we argue that an analogous story can be told in a different physical medium, the brain. Individual neurons and synapses are noisy and unreliable devices; yet behavior, perception, and memory are remarkably robust. Population-coding frameworks and attractor-network models suggest that neural networks in the brain achieve reliability by distributing information over many neurons so that noise averages out~\cite{BurakFiete2009,ChaudhuriFiete2016}. An explicit example is provided by entorhinal grid cells. Sreenivasan and Fiete showed that multi-periodic firings implement an analog error-correcting code that supports path integration with high resolution despite substantial neural noise~\cite{SreenivasanFiete2011}. More recently, Zlokapa et al.~used biological error-correcting codes to construct fault-tolerant artificial neural networks, demonstrating a transition from faulty to reliable computation analogous to fault-tolerance thresholds in QEC~\cite{Zlokapa2024}. These examples demonstrate that error-correcting structure is not unique to engineered digital or quantum devices, but also appears in biological neural systems.

\begin{figure*}[th]
  \centering
  \includegraphics[width=0.9\linewidth]{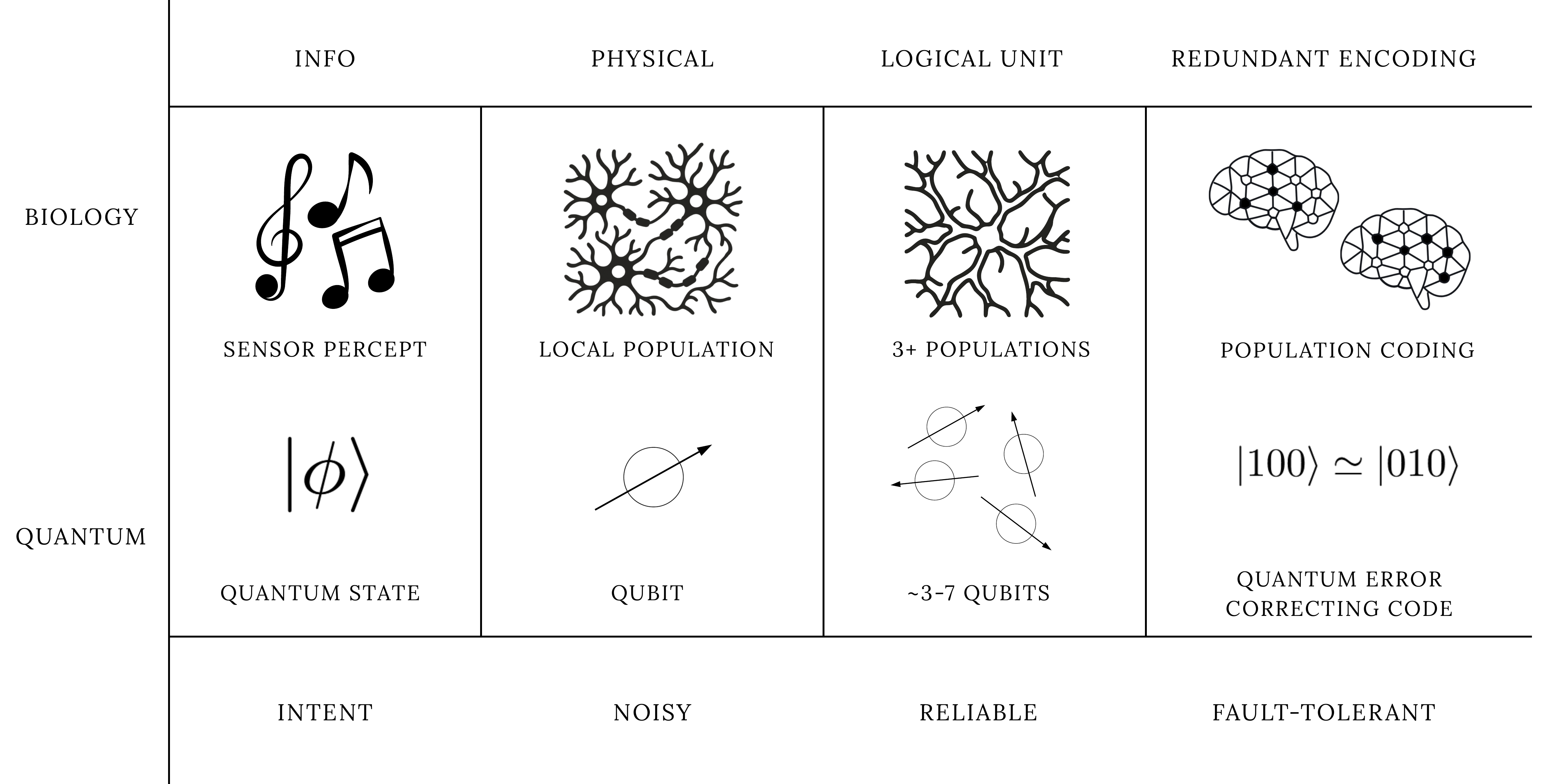}
  \caption{
  An illustration of the structural analogy between QEC and BEC. Each system protects information by moving from noisy physical units to redundant logical representations. In biology, a percept is represented with noisy microcircuits of neurons that are stabilized through population coding to form local neuronal populations. In quantum, a state is represented with noisy physical qubits that are stabilized as code block to form a logical qubit.}
  \label{fig:qec_bec_overview}
\end{figure*}

Crucially, we do \emph{not} assume or invoke quantum coherence or entanglement in biological tissue. While quantum mechanics has been hypothesized to play a role in neural computation and consciousness~\cite{adams2020quantum,neven2024testing}, the hypothesis is controversial and empirical studies have called for further research~\cite{craddock2025quantum,derakhshani2022crossroad,wahbeh2022if}. Instead, we treat neurons as noisy classical elements. Entanglement appears only in the QEC examples. The analogy is therefore between mathematical structures and the organization of observable activity. We are mapping \emph{structure and behavior, not physics}.

Concretely, we define translations between objects in QEC and BEC (see Fig.~\ref{fig:qec_bec_overview}). Logical quantum states map to low-dimensional cognitive variables; physical qubits to local neural populations; stabilizer checks to circuit-level constraints that may include glial modulation; and ancilla qubits to auxiliary interneuron, glial, or collective modes that probe and relay error information. We explore models illustrating these translations. One such model is how neuronal damping can implement syndrome-like error signals and correction in classical networks. We then relate the biological model to experimental observations of bistable population activity, discrete attractor dynamics, and recovery after perturbation, and discuss how this structural correspondence may guide new algorithm and circuit designs. In doing so, we aim to place von Neumann's original question about reliable computation from unreliable parts within a modern framework that spans both quantum devices and the brain.

This perspective suggests a possible transfer from neuroscience to QEC. Neural circuits maintain robust representations using local feedback that operates continuously and adapts as noise statistics change. QEC relies on local, repeated syndrome extraction, but many decoders are designed for fixed or slowly calibrated noise models. Biological error-control mechanisms may therefore suggest adaptive decoders that update online in response to correlated or nonstationary noise while preserving logical information.

Our contribution is not the claim that neural circuits implement QEC, but a minimal stochastic-dynamical construction showing how constraint damping in a neural population code can share the generator structure and steady-state violation scaling of a simple QEC recovery model. By defining a common structural vocabulary across engineered QEC and organic BEC, we aim to clarify what concepts and patterns are universal. Furthermore, we hope that this analogy will help develop improved methods for robust quantum decoding and fault-tolerant architectures.

% \ma{Structure idea:
%   \begin{itemize}
%     \item Small intro on qubits, superposition and entanglement (1 paragraph)
%     \item quantum error correction introduction and explantion
%     \begin{itemize}
%       \item bit flip code
%       \item phase flip code
%       \item shor's code 
%       \item surface code
%     \end{itemize}
%     \item biology terms intro (not enough experience to know what to do here)
%     \item biology error correction intro (not enough experience to know what to do here)
%     \item role of time in our situation (not sure if this should be here or after)
%   \end{itemize}}

\section{Primers on error correction: Quantum and Biological}
\label{sec:primers}

\subsection{Introduction to Quantum Computing}
\label{subsec:qec_primer}

 In quantum computing, a quantum bit, or \emph{qubit}, is the fundamental unit of quantum information. Similar to a classical bit, a qubit has two computational basis states, written in
 \emph{Dirac notation} as $\ket{0}$ and $\ket{1}$. The vertical bar and angle bracket are read together as a \emph{ket}. Unlike a classical bit, a qubit need not be in only one of these two basis states; it can also be in a normalized complex linear combination, or \emph{superposition}, of both:
 \begin{equation}
\ket{\psi}=\alpha\ket{0}+\beta\ket{1}.
\end{equation}
 where $\ket{\psi}$ is a \emph{quantum state}, and $\alpha$ and $\beta$ are the complex amplitudes of $\ket{0}$ and $\ket{1}$, respectively. Quantum states follow a strict set of rules which include:
 \begin{itemize}
  \item \emph{Normalization}: A quantum state is normalized when $|\alpha|^2 + |\beta|^2 = 1$
  \item \emph{Complex Amplitude}: Amplitudes of a quantum state must belong to the complex space where $\alpha, \beta \in \mathbb{C}$
  \item \emph{Measurement postulate}: When a quantum state is measured, it collapses onto the measurement basis. For simplicity, the measurement basis is selected as $\ket0$ and $\ket1$. A measured qubit has probabilities $|\alpha|^2$ and $|\beta|^2$ of collapsing to $\ket{0}$ and $\ket{1}$ respectively.
\end{itemize}

Quantum states can be combined to model a higher dimensional space, represented algebraically with a \emph{tensor product}:
\begin{equation}
  \ket{\psi}=\ket{\psi_A} \otimes \ket{\psi_B}
  %\\
  %\ket{01}=\ket0\otimes\ket1
\end{equation}
States that can be factorized into individual qubit states are called \emph{product states}. States that cannot be written in this product form are \emph{entangled states}; for such states, measurement outcomes can exhibit correlations that are not reducible to independent single-qubit states.
 
One example of an entangled state is presented by the \emph{Bell state}:
\begin{equation}
  \ket{\Phi^+} = \frac{1}{\sqrt{2}} \left( \ket{00} + \ket{11} \right).
\end{equation}
For this state, measurements in the computational basis are perfectly correlated. If the first qubit is measured as $\ket{0}$, the post-measurement state is $\ket{00}$; if it is measured as $\ket{1}$, the post-measurement state is $\ket{11}$.

\subsection{Quantum error correction}

QEC asks how one can protect an unknown quantum state from noise without directly measuring that state. There are fundamental differences to classical error correction that make QEC subtler and more challenging. A classical bit can be copied, compared to other copies, and majority-voted if one copy is corrupted. An unknown qubit cannot be cloned, and a direct measurement of the qubit generally collapses the superposition. QEC therefore protects information indirectly: it encodes one \emph{logical qubit} into a larger collection of noisy \emph{physical qubits} such that likely errors can be detected and corrected without revealing the logical state itself~\cite{Shor1995,NielsenChuang2010,Terhal2015}.

Consider a basic quantum state
\begin{equation}
\ket{\psi}=\alpha\ket{0}+\beta\ket{1}.
\end{equation}
Instead of storing this state in a single physical qubit, we store
\begin{equation}
\ket{\psi_L}=\alpha \ket{0_L}+\beta \ket{1_L} ,
\end{equation}
where $\ket{0_L}$ and $\ket{1_L}$ are logical qubits, or \emph{codewords}. The set of allowed encoded states forms the \emph{codespace} $\mathcal{C}\uq$. The central idea is that typical local noise perturbs the physical qubits in a way that can be detected at the level of collective constraints, before it becomes an error on the logical information itself.

A simple example is the three-qubit bit-flip code,
\begin{equation}\label{eqn:three_qubit_repetition}
\ket{0_L}=\ket{000},\qquad \ket{1_L}=\ket{111}.
\end{equation}
 Note that this encoding does not copy the unknown state $\ket{\psi}$. Instead, it stores the logical information in correlations among physical qubits. The basis codewords $\ket{000}$
and $\ket{111}$ are product states but a generic encoded superposition $\alpha\ket{000}+\beta\ket{111}$ is entangled. With this representation, if one physical qubit flips, then the encoded state leaves the set of consistent patterns. One can detect which qubit flipped by measuring the parities $Z_1Z_2$ and $Z_2Z_3$. These measurements do \emph{not} ask whether the logical state is $\ket{0_L}$ or $\ket{1_L}$; instead, they ask whether neighboring qubits agree. Thus they extract \emph{error information} without revealing the amplitudes $\alpha$ and $\beta$. This example illustrates the basic architecture of QEC, although it protects only against bit-flip errors. Full QEC must also protect against \emph{phase} errors, which have no classical analog. Shor's original code and the more general stabilizer formalism extend the same logic to arbitrary single-qubit errors~\cite{Shor1995,NielsenChuang2010,Terhal2015}.

The stabilizer picture is especially useful for the structural comparison developed later in this paper. In a stabilizer code, one specifies a set of commuting check operators $S_j\uq$ whose simultaneous $+1$ eigenspace defines the codespace:
\begin{equation}
\mathcal{C}\uq
=
\left\{
\ket{\psi} : S_j\uq\ket{\psi}=\ket{\psi}
\text{ for all } j
\right\}.
\end{equation}
Each $S_j\uq$ is a \emph{constraint} on the joint state of several physical qubits. If an error occurs, one or more of these constraints may be violated. Measuring the checks produces a set of binary outcomes
\begin{equation}
\bm{s}\uq=(s_1\uq,\dots,s_m\uq),\qquad s_j\uq\in\{\pm1\},
\end{equation}
called the \emph{error syndrome}. A value $s_j\uq=-1$ indicates that the corresponding constraint has been violated.

In practice, these checks are not measured directly on the data qubits. Instead, the data qubits $q_i\uq$ are coupled to auxiliary \emph{ancilla qubits} $a_k\uq$ prepared in known states. The ancillas interact with the data through a syndrome-extraction circuit and are then measured. In this way, the ancillas act as temporary carriers of syndrome information. The protected logical state remains encoded nonlocally in the data qubits, while the ancillas report which checks were satisfied or violated. This separation between \emph{protected information} and \emph{measured information} is one of the defining ideas of QEC.

A second important point, which is often counterintuitive at first, is that QEC can correct continuous noise using discrete syndrome data. A generic perturbation of a qubit can change amplitudes and phases continuously. However, once one measures the stabilizer checks, the relevant information is represented in a discrete syndrome. In the standard qubit language, arbitrary local errors can be decomposed into bit flips ($X$), phase flips ($Z$), and combinations. QEC does not need to reconstruct the full error; it needs just enough information to decide which recovery operation is most likely to restore the logical state.

The \emph{decoder} is the map that makes this decision. Given a syndrome pattern, or more realistically a time history of syndrome patterns from repeated measurements,
the decoder $\mathcal{D}\uq$ outputs a \emph{recovery operation} $\mathcal{R}\uq$:
\begin{equation}
\mathcal{D}\uq:\bm{s}\uq\mapsto \mathcal{R}\uq .
\end{equation}
Importantly, the decoder is usually a classical algorithm, even though the information being protected is quantum. A logical error occurs when the physical noise and recovery together enact a nontrivial transformation on the encoded information.

For the structural comparison below, the important point is not any particular code construction, but the organization of the correction process: noisy physical degrees of freedom support protected logical variables, constraints define a codespace, syndrome extraction reports constraint violations without measuring the logical state, and a decoder selects a recovery. In fault-tolerant architectures this cycle is repeated using imperfect hardware, and below a threshold physical error rate, increasing the code size can reduce the logical error rate~\cite{NielsenChuang2010,Terhal2015}. This combination of redundant encoding, repeated constraint checking, syndrome-based inference, and recovery is the organizational pattern that we compare to error control in neural systems.

\subsection{Neurons and Computation}

Neurons are the central computational units within the brain. They receive electrical signals through dendrites, which form synapses with the axons of other neurons. When a neuron’s membrane potential exceeds threshold, it fires an action potential that propagates along its axons, potentially triggering connected neurons to propagate the signal.

Compared to digital computing, neurons compute fundamentally differently. They are fully asynchronous, with each neuron separately reacting to incoming electrical signals. Neurons operate in both the analog and discrete domains, performing calculations with varying voltages but sending information as discrete spiking events.

\subsubsection{Neural population coding and biological error correction}

Neurons encode information, including sensory stimuli, through a phenomenon known as population coding~\cite{JOHNSON2000563}. A significant amount of prior work has focused on how stimuli are encoded through population coding, which is essential for understanding how neurons represent complex, high-dimensional signals through spiking dynamics~\cite{Li_2024, Safaai2025, Xie2024}. However, population coding is also essential for error correction.

Tka\v{c}ik et al. provide an overview of population coding-based error correction through redundant information, finding that, as noise and stimulus changed, the optimal coding and amount of redundant information did as well~\cite{doi:10.1073/pnas.1004906107}. Boerlin et al. and Bourdoukan et al. argue that the coding communicates the error level to the system, which allows for the system to self-correct by integrating the signals from other neurons above an error threshold~\cite{Boerlin2013, NIPS2012_3a15c7d0}. Both Berry and Tka\v{c}ik propose that neural activity patterns can be clustered to represent population codewords~\cite{10.3389/fncom.2020.00020}. These clusters are error correcting despite the noise associated with individual neurons.

In addition to proposing the idea of population codeword clusters, Berry and Tka\v{c}ik discuss how the dynamics of the neuronal system affect population coding and error rates, finding that the system resides in a subcritical, glassy state~\cite{10.3389/fncom.2020.00020}. Calvo et al. support this, also finding that the brain operates in a subcritical system while remaining close to criticality~\cite{36v9-wtm8}. They argue that this allows for a robust system with a long lifetime. This echoes findings on reservoir computing, where being close to criticality is seen as essential for high performance~\cite{10.1098/rstb.2018.0377}. Calaim et al. discuss the distance to criticality, using a spiking model to draw a `bounding box' regime where the network is robust~\cite{10.7554/eLife.73276}.

Other work has examined robustness that stems from the connectivity and circuitry of the neuronal network. Arle et al. use a spiking neuron model to simulate the brain, and find that more complex, connected systems, such as the brain, are inherently more stable and robust~\cite{Arle2020}. Lim and Goldman analyze the microcircuitry of the brain, finding that it generates a negative corrective signal that compensates for errors caused by noisy-firing neurons, allowing them to develop a model of short-term memory that is similarly robust to the brain~\cite{Lim2013}.

\subsubsection{Individual neuron complexity and population coding}

Neurons are significantly more complex than qubits, a feature that makes them less analogous than this comparison might suggest. Beniaguev et al. show that modeling a single neuron, with its N-methyl-D-aspartate (NMDA) receptors, requires a 5 hidden layer artificial neural network, while a neuron without its NMDA receptors requires only 1 hidden layer~\cite{BENIAGUEV20212727}. Similarly, Aizenbud et al. show that a single neuron is capable of solving complex problems, like XOR, through its dendritic networks~\cite{Aizenbud2026.06.08.730984}. Findings like these have inspired the field of dendritic computing.

In contrast, qubits are closer to classical bits, where they are basic units of data and are not capable of compute by themselves. Therefore, the capability gap between neurons and qubits calls into question the accuracy of this metaphor. Despite their complexity, neuronal units can be, and are, usefully approximated as simple binary variables in reduced models and rely on population coding to develop complex representations and suppress errors~\cite{Li_2024, Safaai2025, Xie2024, doi:10.1073/pnas.1004906107}. Accordingly, the model employed in Sec.~\ref{sec:bec_numeric_experiment} is a reduced, excitatory-only population rate model of the Wilson–Cowan type~\cite{WILSON19721, ErmentroutTerman2010_ch11_12}.

\section{Structural dictionary between quantum error correction and neural population codes}
\label{sec:dictionary}

%\ws{we are defining the terms in a dictionary but we are also providing a mapping, so i think translation-dictionary (thesaurus++ ?) is a better description.}

In this section we introduce a structural dictionary to translate between QEC and neural population codes, with the goal of preparing a quantitative, mathematical treatment in Sec.~\ref{sec:quantitative}. 
%We assume that the reader is broadly familiar with stabilizer-based QEC schemes~\cite{NielsenChuang2010,Terhal2015} and with population codes and attractor-network models in neuroscience~\cite{BurakFiete2009,ChaudhuriFiete2016,SreenivasanFiete2011}.
% ws & ma : perhaps these readings should be cited where used in the definitions
Our aim is to identify which quantum objects map to which biological objects, and specify what aspects of their structure can be exploited.

We use a superscript $(Q)$ for quantities associated with a quantum code and a superscript $(B)$ for the corresponding biological implementation. For example, $\mathcal{C}\uq$ denotes a QEC code space and $\mathcal{M}\ub$ a neural manifold. The mapping we propose is \emph{structural}: we treat neurons as classical stochastic dynamical elements, and we do not assume quantum coherence or entanglement in the brain. The analogy is between roles rather than physical phenomena, with shared roles including redundant encoding, constraint checks, syndrome-like signals, and decoding.

We organize the dictionary into four groups of objects:

\begin{enumerate}
    \item core structural elements, presented in Sec.~\ref{subsec:core_objects},
    \item checks, syndromes, and ancillary degrees of freedom, presented in Sec.~\ref{subsec:checks_syndromes},
    \item decoding and correction dynamics, presented in Sec.~\ref{subsec:decoding_correction}, and
    \item design, evolution, and thresholds, presented in Sec.~\ref{subsec:design_thresholds}.
\end{enumerate}

\subsection{Core structural objects}
\label{subsec:core_objects}

The first group of objects concerns what information is represented, where that information is encoded, how noise acts on the underlying physical degrees of freedom, and how robustness is quantified.

On the QEC side, we take as a reference point a stabilizer code embedded in a larger fault-tolerant architecture~\cite{NielsenChuang2010,Terhal2015}. The basic ingredients are:
\begin{itemize}
 \item \emph{Physical qubits} $q_i\uq$ that are subject to noise.
 \item \emph{Logical qubits} $Q_\alpha\uq$ (or logical states) encoded in an error-correcting code.
 \item A \emph{code space} $\mathcal{C}\uq \subset \mathcal{H}^{\otimes n}$, defined as the simultaneous $+1$ eigenspace of a commuting group of stabilizer operators.
 \item A \emph{noise channel} $\mathcal{N}\uq$ acting on the physical degrees of freedom.
 \item A \emph{code distance} $d\uq$ that characterizes how many local errors are required to cause a logical error.
\end{itemize}

On the biological side, we consider a recurrently connected neural population, as in the experimental and theoretical work reviewed in~\cite{BurakFiete2009,ChaudhuriFiete2016,Santello2019,KofujiAraque2021,ONeill2023,Lee2014,Murphy-Royal_Ching_Papouin_2023}. Here the natural counterparts are:
\begin{itemize}
 \item \emph{Physical units} $x_i\ub$, which may be individual neurons, synapses, or small microcircuits, each with noisy, variable response properties. Sec.~\ref{sec:bec_numeric_experiment} instantiates $x_i\ub$ as a small local population.
 \item \emph{Low-dimensional variables} $X_\alpha\ub$ representing quantities encoded by the population code including spatial position, working-memory state, and muscle-control signals. 
 \item A \emph{neural manifold} or \emph{attractor set} $\mathcal{M}\ub \subset \mathbb{R}^N$ (or a suitable state space) consisting of patterns of activity across neurons that implement particular values of the low-dimensional variables. This is downstream from the connectome, which is the physical wiring of the neurons in the brain.
 \item An effective \emph{noise process} $\mathcal{N}\ub$ with fast neural variability (spiking noise, synaptic failures). This does not include slow, structural changes like neuroplasticity, which plays a role in learning but does not cause the type of errors we are interested in.
 \item A notion of \emph{robustness scale} $d\ub$ quantifying the size of perturbations in the physical units (e.g.\ fraction of neurons or synapses perturbed) that can be tolerated before the encoded variable $X_\alpha\ub$ changes appreciably. Computational experiments studying how perturbations can push the brain to criticality quantify this notion~\cite{10.7554/eLife.73276, Arle2020, 36v9-wtm8}.
 
\end{itemize}

Thus the first layer of the dictionary is:
\begin{equation}
 \bigl\{ q_i, Q_\alpha, \mathcal{C}, \mathcal{N}, d \bigr\}\uq
  \ \longleftrightarrow\
 \bigl\{ x_i, X_\alpha, \mathcal{M}, \mathcal{N}, d \bigr\}\ub.
\end{equation}
In Sec.~\ref{sec:quantitative} we will use this correspondence to define constraint-violation observables and to compare $V_\infty\uq$ and $V_\infty\ub$ on equal footing for concrete models.

\subsection{Checks, syndromes, and ancillary degrees of freedom}
\label{subsec:checks_syndromes}

The second group of objects concerns how errors are \emph{detected}. In stabilizer QEC, this is achieved by measuring a set of commuting operators $S_j\uq$ (stabilizer generators) on the physical qubits. The outcome of each measurement is a binary \emph{syndrome bit} $s_j\uq \in \{\pm 1\}$ indicating whether the corresponding stabilizer constraint is satisfied. In many schemes, these measurements are implemented by coupling the data qubits $q_i\uq$ to ancillary qubits $a_k\uq$, which are prepared in a known state, entangled with the data via a syndrome-extraction circuit, and then measured in the computational basis~\cite{NielsenChuang2010,Terhal2015}. The ancillas thus serve as dedicated carriers of syndrome information.

In neural systems, there is no exact analog of a projective stabilizer measurement, but several layers of circuitry implement \emph{constraints} and \emph{mismatch signals} that play related roles. Recurrent inhibitory and excitatory loops enforce consistency conditions on population activity~\cite{BurakFiete2009,ChaudhuriFiete2016,SreenivasanFiete2011}. Astrocytes may add an additional, slower layer of monitoring and modulation~\cite{Santello2019,KofujiAraque2021,ONeill2023,Lee2014,Murphy-Royal_Ching_Papouin_2023}, however that is not the focus of this analogy (see Sec.~\ref{sec:fw} for a discussion of future work along these lines).

We identify the following structural correspondence: 
\begin{itemize} 
 \item \emph{Stabilizer checks} $S_j\uq$ may map to \emph{circuit-level constraints} $C_j\ub$ on joint activity patterns, such as consistency of firing with a learned manifold $\mathcal{M}\ub$, maintenance of excitation--inhibition balance, or preservation of particular phase relationships in ongoing oscillations, with possible astrocyte-mediated contributions. 
 \item \emph{Syndrome bits} $s_j\uq$ map to \emph{mismatch signals} $m_j\ub$, which can be realized by deviations of neural activity from $\mathcal{M}\ub$, by prediction errors in circuits that implement internal models signaling that the current pattern of activity is out of learned constraints. 
 \item \emph{Ancilla qubits} $a_k\uq$ map to \emph{ancillary degrees of freedom} $a_k\ub$ that monitor and relay information about the state of the network without directly encoding the low-dimensional variables $X_\alpha\ub$. These could include interneurons, collective oscillatory modes, and possibly astrocytes that are more sensitive to constraint violations than to the exact value of the encoded variable.
\end{itemize}

In this view, interneurons, collective oscillatory modes, and possibly astrocytes are not quantum ancillas, but they can play an ancillary \emph{role}: they may carry syndrome-like information about whether local neural activity and environmental conditions are compatible with the current code manifold $\mathcal{M}\ub$. 

\subsection{Decoding and correction dynamics}
\label{subsec:decoding_correction}

The third group of objects concerns how detected errors are \emph{corrected}. On the quantum side, a \emph{decoder} is an algorithm (or, more generally, a map) that takes a syndrome pattern $(s_j\uq)$ as input and outputs a recovery operation $\mathcal{R}\uq$ acting on the data qubits. In stabilizer codes, $\mathcal{R}\uq$ is typically chosen from the Pauli group, and the decoder is designed to minimize the probability of a logical error given a noise model~\cite{Terhal2015}. In fault-tolerant architectures, such decoders are applied repeatedly in time to keep the logical state within the code space $\mathcal{C}\uq$.

In neural population codes, much of the ``decoding'' and ``correction'' is implemented by the intrinsic dynamics of the network. Prominent examples are attractor models of memory: starting from a noisy pattern of activity, interactions drive the system toward a family of stable states (an attractor manifold), thus correcting microscopic noise and preserving the encoded variable~\cite{BurakFiete2009,ChaudhuriFiete2016,SreenivasanFiete2011}.

We identify the following structural correspondence:
\begin{itemize} 
 \item A \emph{decoder map} $\mathcal{D}\uq : (s_j\uq) \mapsto \mathcal{R}\uq$ on the QEC side with an effective \emph{error-control map} $\mathcal{D}\ub : (m_j\ub) \mapsto \Delta \theta\ub$ on the biological side, where $\Delta \theta\ub$ denotes changes in neuronal and astrocytic parameters (e.g.\ synaptic weights, intrinsic excitability). 
 \item The \emph{recovery operation} $\mathcal{R}\uq$ corresponds to fast \emph{relaxation dynamics} toward an attractor manifold $\mathcal{M}\ub$ (correcting noise).
\end{itemize}

 An example of the recovery operation within neuroscience is provided by Nakazawa et al., who find that the \emph{CA3} region of the hippocampus can rebuild a full memory from a partial or noisy cue~\cite{doi:10.1126/science.1071795}. Additionally, physical evidence for $\mathcal{M}\ub$ was found by Kim et al., who found that a \emph{Drosophila}'s brain exhibited ring attractor dynamics after optogenetic displacement of the activity bump, consistent with the attractor manifold we propose~\cite{doi:10.1126/science.aal4835}. Finally, from computational studies, the glia-inspired rhythmic-sharing algorithms provide an artificial example of continuous syndrome-like monitoring: oscillatory modulation of network couplings introduces ancillary degrees of freedom, and deviations of synchrony or phase-locking structure from a learned baseline act as mismatch signals. The phase dynamics can then reconfigure effective information pathways when input statistics change~\cite{WhitehouseKangLosert2025}. This is a classical, continuous analog of constraint monitoring coupled to adaptive recovery. In Sec.~\ref{sec:quantitative} we formalize $\mathcal{R}\ub$ in a tractable model and quantify its error-suppressing properties.
 
\subsection{Design, evolution, and thresholds}
\label{subsec:design_thresholds}

Finally, we consider how codes and circuits are \emph{designed}, and what notions of \emph{threshold} apply. In the QEC setting, codes are engineered to optimize distance and locality, subject to hardware and architectural constraints~\cite{Terhal2015}. Threshold theorems then guarantee that if the physical error rate is below a critical value $p_{\mathrm{th}}\uq$, and if sufficient resources are allocated, accurate logical computation is possible.

Biological systems, by contrast, are shaped by evolution, development, and learning, and not engineered with an explicit code in mind. However, as we argue in this work, numerous works show they have code-like structure: redundantly encoded variables, clustered population activity, and error-correcting dynamics that can be analyzed in the language of codes and decoders. Like QEC settings, neural circuits are constrained by their ``hardware'' and optimized for distance and locality~\cite{Barranca2026-xp}. Recent work has shown that biologically inspired error-correcting codes can be used to construct fault-tolerant artificial neural networks, with a transition between reliable and unreliable computation as the rate of neuron failure is varied, similar to QEC thresholds~\cite{Zlokapa2024}.

% Furthermore, like quantum control systems, they have controllers, the astrocytes, which play an understudied role in structuring the neuronal system. 

\begingroup
\setlength{\tabcolsep}{6pt}
\renewcommand{\arraystretch}{1.2}
\small
\begin{table*}[t]
 \centering
 \caption{A structural dictionary between the core objects of quantum error correction (QEC) and neural population codes. Superscripts $(Q)$ and $(B)$ denote quantum and biological quantities, respectively.}
 \label{tab:dictionary_core}
 \begin{tabularx}{\textwidth}{l X X}
  \toprule
  QEC element & Biological counterpart & Informal role \\ \midrule
  Physical qubits $q_i\uq$ &
   Physical units $x_i\ub$ (neuronal microcircuits \& local populations) &
   Noisy elements that directly experience the noise process. \\
  Logical qubits $Q_\alpha\uq$ &
   Low-dimensional variables $X_\alpha\ub$ (cognitive variables) &
   Information to be protected and made behaviorally robust. \\
  Code space $\mathcal{C}\uq$ &
   Neural manifold / attractor set $\mathcal{M}\ub$ &
   Set of valid global states implementing particular values of
   $X_\alpha\ub$. \\
  Noise channel $\mathcal{N}\uq$ &
   Neural noise and drift process $\mathcal{N}\ub$ &
   Perturbations due to spiking variability, synaptic failures,
   neuromodulation, and slow nonstationarities. \\
  Code distance $d\uq$ &
   Robustness scale $d\ub$ &
   Tolerable size of perturbation in physical units before the encoded
   variable changes appreciably. \\
  \bottomrule
 \end{tabularx}
\end{table*}
\endgroup

\section{Quantitative framework for biological and quantum error correction}
\label{sec:quantitative}

We now construct a quantitative model for the structural dictionary of Table~\ref{tab:dictionary_core}. Our goal is to describe both QEC and biological error correction as \emph{redundant encoding plus constraint-based inference}. On the quantum side, noisy physical qubits $q_i\uq$ support protected logical variables $Q_\alpha\uq$ inside a code space $C\uq$. On the biological side, noisy physical units $x_i\ub$ (neurons, synapses, local populations, or small microcircuits) support protected low-dimensional variables $X_\alpha\ub$ inside a neural manifold or attractor set $M\ub$. The analogy is structural, organized around distributed redundancy, constraint checks, syndrome-like variables, and recovery dynamics.

These ideas appear in many other areas of physics. We name a few familiar examples. In thermodynamics, a given macrostate corresponds to many microstates, and is therefore protected against ``noise'' that changes microstates of individual particles (atoms). In general relativity, the Einstein equations can be formulated as a system of hyperbolic evolution equations that preserve constraints. Constraint damping or projection methods bring the evolutionary dynamics back to the constraint manifold.

\subsection{Quantum}

For a stabilizer code, the protected quantum states lie in a code manifold that we write as
\begin{equation}
C\uq =
\left\{
|\psi\rangle \in \mathcal{H}^{\otimes n}
:
S_j\uq |\psi\rangle = |\psi\rangle,\;\; j=1,\dots,m
\right\},
\label{eq:qec_codespace}
\end{equation}
where the commuting operators $S_j\uq$ are the stabilizer checks.
Under a noisy channel $N\uq$, syndrome extraction leads to a binary
syndrome vector
\begin{equation}
s\uq(t) = \left(s_1\uq(t),\dots,s_m\uq(t)\right),
\qquad
s_j\uq(t)\in\{\pm 1\},
\label{eq:qec_syndrome}
\end{equation}
and a decoder $D\uq$ maps the syndrome history to a recovery
operation $R_t\uq$,
\begin{equation}
R_t\uq = D\uq\!\left(s\uq(0{:}t)\right).
\label{eq:qec_decoder}
\end{equation}
A full correction cycle can therefore be written schematically as
\begin{equation}
\rho_{t+1} = R_t\uq \circ N_t\uq(\rho_t).
\label{eq:qec_cycle}
\end{equation}
The logical error rate decreases with code distance $d\uq$
whenever the physical noise is below threshold:
\begin{equation}
p_L\uq(d\uq,p)
\approx
A_Q
\left(
\frac{p}{p_{\mathrm{th}}\uq}
\right)^{(d\uq+1)/2},
\qquad
p < p_{\mathrm{th}}\uq.
\label{eq:qec_threshold_scaling}
\end{equation}

\subsection{Numerical implementation of a QEC experiment}
\label{sec:qec_numerical_experiment}

To make the structural dictionary quantitative, we perform a numerical experiment where both the quantum and biological systems are treated as noisy dynamics on a larger state space together with a correction mechanism that steers trajectories to a protected manifold. On the quantum side this protected set is the codespace \(C\uq\); on the biological side it is the neural code manifold \(M\ub\). The same class of observables is then compared across the two systems: an encoded-information observable and a manifold-violation observable.

As a minimal QEC model we use the three-qubit repetition code \eqref{eqn:three_qubit_repetition}, which protects against bit-flip errors~\cite{Shor1995,Gottesman1997,Terhal2015}. The logical codewords are
\begin{equation}
|0_L\rangle = |000\rangle,
\qquad
|1_L\rangle = |111\rangle,
\label{eq:qec_rep_logical_basis}
\end{equation}
and the stabilizer checks are
\begin{equation}
S_1\uq = Z_1 Z_2,
\qquad
S_2\uq = Z_2 Z_3.
\label{eq:qec_rep_stabilizers}
\end{equation}
For a generic encoded state
\begin{equation}
|\psi_L\rangle = \alpha |0_L\rangle + \beta |1_L\rangle,
\qquad
|\alpha|^2+|\beta|^2=1,
\label{eq:qec_rep_generic_state}
\end{equation}
we evolve the density operator \(\rho\) under independent bit-flip noise and compare discrete and continuous recovery. In the numerical implementation we choose a generic superposition with non-vanishing amplitudes to demonstrate decay of multi-qubit errors.

The discrete-time bit-flip channel is
\begin{equation}\label{eq:qec_rep_discrete_noise}
\mathcal N_p\uq(\rho) = \sum_{e\in\{0,1\}^3} p^{|e|}(1-p)^{3-|e|} X(e)\,\rho\,X(e)^\dagger,
\end{equation}
where
\begin{equation}
X(e)=X_1^{e_1}X_2^{e_2}X_3^{e_3},
\qquad
|e|=e_1+e_2+e_3.
\end{equation}
Syndrome sectors are resolved by the projectors
\begin{equation}\label{eq:qec_rep_projectors}
P_{s_1,s_2} = \frac{1}{4} \left(I^{\otimes 3}+s_1 S_1\uq\right) \left(I^{\otimes 3}+s_2 S_2\uq\right),
\end{equation}
with $s_1,s_2\in\{\pm 1\}$, and correction operators
\begin{equation}\label{eq:qec_rep_recovery_ops}
R_{++}=I, \quad
R_{-+}=X_1, \quad
R_{--}=X_2, \quad
R_{+-}=X_3.
\end{equation}
The corresponding syndrome-sector recovery map is
\begin{equation}
\mathcal R\uq(\rho)
=
\sum_{s_1,s_2}
R_{s_1,s_2}\,
P_{s_1,s_2}\,
\rho\,
P_{s_1,s_2}\,
R_{s_1,s_2}^\dagger.
\label{eq:qec_rep_recovery_map}
\end{equation}
A discrete correction cycle is then
\begin{equation}
\rho_{n+1}
=
\mathcal R\uq\!\circ \mathcal N_p\uq(\rho_n).
\label{eq:qec_rep_discrete_cycle}
\end{equation}

To compare projections and damping, we evolve the system in continuous time according to
\begin{equation}\label{eq:qec_rep_continuous_generator}
\frac{d\rho}{dt} = \gamma \sum_{i=1}^3 \left(X_i \rho X_i - \rho\right) +
\kappa\left(\mathcal R\uq(\rho)-\rho\right),
\end{equation}
where \(\gamma\) is the physical bit-flip rate and \(\kappa\) is the recovery rate. The first term continuously injects syndrome violations, while the second term continuously steers the state back to the codespace. Numerically, the discrete model is iterated by direct application of the channel and recovery map, while the continuous generator is integrated with a fourth-order Runge--Kutta scheme. Continuous recovery generators of this form follow ~\cite{AhnDohertyLandahl2002,SarovarMilburn2005,KerckhoffEtAl2010}. 

\begin{figure*}[ht]
\centering
\includegraphics[width=\textwidth]{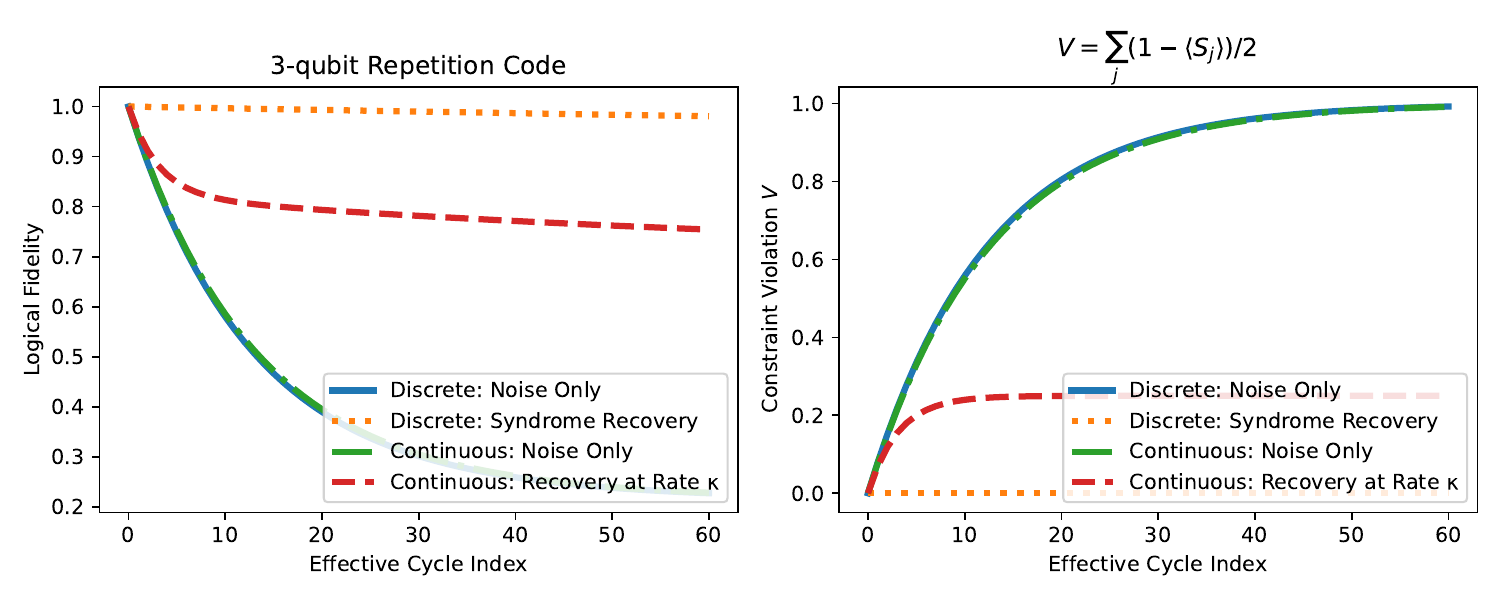}
\caption{
Numerical simulation of the three-qubit repetition code comparing discrete and continuous recovery. Left: logical fidelity \(F_L\uq(t)\) given in \eqref{eq:qec_rep_fidelity}.
Right: stabilizer violation \(V\uq(t)\) given in \eqref{eq:qec_rep_constraint_violation}. The blue solid curve shows discrete-time bit-flip noise without recovery; the orange solid curve shows discrete syndrome recovery after each cycle; the green dashed curve shows continuous-time bit-flip noise without recovery; and the red dashed curve shows continuous-time recovery at finite rate \(\kappa\). Exact syndrome recovery returns the state to the codespace after each cycle, so \(V\uq=0\) identically, while the logical fidelity still decays due to uncorrectable multi-qubit errors. Continuous-time recovery suppresses stabilizer violation but leaves a nonzero steady-state residual under persistent noise, demonstrating the difference between projection recovery and damping recovery.
}
\label{fig:qec_repetition_demo}
\end{figure*}

The quantum observables recorded in the simulation are
\begin{align}
F_L\uq(t) &= \langle \psi_L|\rho(t)|\psi_L\rangle, \label{eq:qec_rep_fidelity} \\
P_C\uq(t) &= \mathrm{Tr}\!\left(P_{++}\rho(t)\right), \label{eq:qec_rep_codespace_population} \\
V\uq(t) &= \sum_{j=1}^2 \frac{1-\mathrm{Tr}\!\left(S_j\uq\rho(t)\right)}{2}, \label{eq:qec_rep_constraint_violation}
\end{align}
where \(F_L\uq\) is the logical fidelity, \(P_C\uq\) the codespace population, and \(V\uq\) a stabilizer violation. For the continuous model, starting from the codespace, we find
\begin{equation}
V_\infty\uq=\frac{4\gamma}{\kappa+4\gamma},
\label{eq:qec_rep_v_infty}
\end{equation}
showing that continuous recovery suppresses, but does not eliminate, off-manifold error under persistent noise. By contrast, the ideal discrete recovery map of Eq.~\eqref{eq:qec_rep_recovery_map} returns the state exactly to the codespace after each correction cycle, although logical fidelity can still decay due to uncorrectable multi-qubit errors.

For the data shown in Fig.~\ref{fig:qec_repetition_demo}, we choose a generic logical superposition with nonvanishing amplitudes and evolve it for \(60\) effective correction cycles. In the discrete model, each physical qubit undergoes an independent bit flip with probability \(p=0.02\) per cycle. In the continuous model, we use \(\gamma=1\), \(\kappa=12\), and \(\Delta t=0.02\), and plot the continuous trajectories against the effective cycle index \(t/\Delta t\) for direct comparison with the discrete-time evolution.

The left panel in Fig.~\ref{fig:qec_repetition_demo} displays the logical fidelity \eqref{eq:qec_rep_fidelity}, while the right panel displays the stabilizer violation \eqref{eq:qec_rep_constraint_violation}. The discrete and continuous noise curves lie nearly on top of each other, showing that the continuous-time generator reproduces the same qualitative degradation of the encoded state as repeated discrete bit-flip noise. Under exact discrete syndrome recovery, \(V\uq(t)=0\) throughout, because each correction cycle returns the state to the codespace exactly. Nevertheless, the logical fidelity still decays slowly, demonstrating uncorrectable multi-qubit bit flips that act nontrivially within the logical subspace. By contrast, continuous recovery at finite rate \(\kappa\) substantially suppresses the growth of stabilizer violation and improves the logical fidelity relative to noise-only evolution, but it does not restore the state exactly to the codespace under persistent noise. Instead, the stabilizer-violation observable approaches the nonzero asymptotic value given in Eq.~\eqref{eq:qec_rep_v_infty}. This explicitly distinguishes recovery due to projection from recovery due to damping: projection enforces exact return to the code manifold after each cycle, whereas damping results in a stationary balance between noise injection and restorative flow.

\subsection{BEC equivalent of the QEC numerical experiment}
\label{sec:bec_numeric_experiment}

\begin{figure*}[ht]
    \centering
    \includegraphics[width=\textwidth]{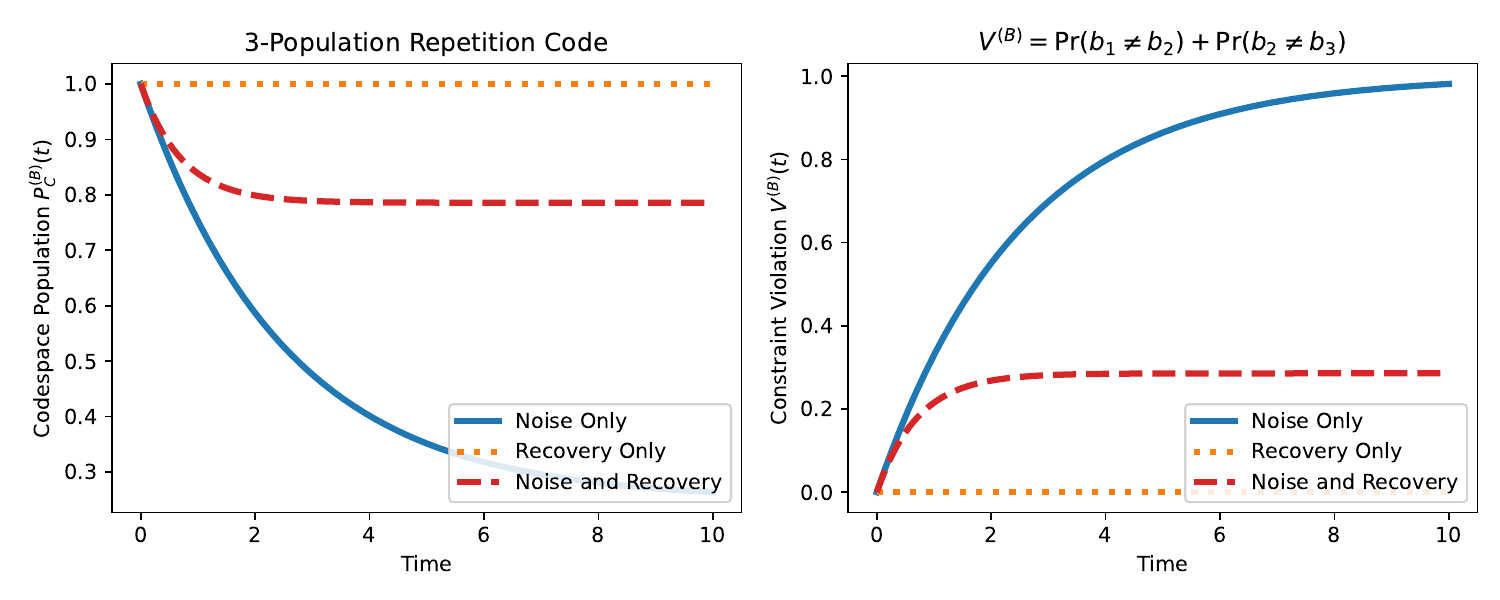}
    \caption{
    The biological counterpart to Fig.~\ref{fig:qec_repetition_demo}: a numerical simulation of a three-population code. Left: codespace population $P_C\ub(t)$.
    Right: constraint violation Eq.~(\ref{eq:bio_constraint}). The blue curve shows noise without recovery; the orange curve shows recovery without noise; and the red dashed curve shows both together. Without recovery, $V\ub$ grows towards its maximum as noise drives units from the two codewords. Recovery alone holds the system exactly on the codespace. With both, continuous recovery suppresses constraint violation but leaves a nonzero steady-state residual. Unlike Fig.~\ref{fig:qec_repetition_demo}, the biological model is integrated in continuous time only. Parameters: $\gamma = 0.1$, $\kappa = 1$, initial condition $b = 000$; Eq.~(36) is integrated with an implicit Radau method~\cite{hairer} (relative and absolute tolerances $10^{-8}$ and $10^{-10}$).
    }
    \label{fig:bec_repetition_demo}
\end{figure*}

A structurally similar reduced model can be constructed for a biological setting, considering that, in the cortex, bistable persistent activity is typically a property of small recurrently connected populations~\cite{Inagaki2019-fe, Sanchez-Vives2000}. Ermentrout and Terman~\cite[Sec.~12.1.3]{ErmentroutTerman2010_ch11_12} present rate models that, although derived from synaptically coupled neurons, describe the mean activity of a local population when its neurons fire approximately independently. We adopt this population interpretation throughout, so that each unit $x_i^{(B)}$ is a recurrently connected population rather than a single neuron. Starting from~\cite[Eq.~(12.6)]{ErmentroutTerman2010_ch11_12}:

\begin{equation}
    \tau_{m,i}\frac{dv_i}{dt} + v_i = \sum_j w_{ij} F_j \left (v_j \left ( t -r \right ) \right )
    \label{eq:bio_et}
\end{equation}

\noindent where $v_i$ is the population's mean somatic potential, which drives firing. We define three identical populations $i=1,2,3$, with a common time constant $\tau$ ($\tau_{m,i} = \tau\ \forall i$) and gain function $\phi$ (hence $F_i = \phi\ \forall i$). We set $r=0$, since recovery is slow compared to axonal delay. We separate the population's recurrent excitation $w_\text{self}$ from the excitation $\wc$ from other populations and add a constant drive $I$, which represents input from outside the three-unit circuit, to get:

\begin{equation}
    \tau\frac{dv_i}{dt} + v_i =  I +w_{\text{self}}\phi(v_i) + \wc\sum_{j \neq i}\phi(v_j)
\end{equation}

Ermentrout and Terman's equation~\cite[Eq.~(12.6)]{ErmentroutTerman2010_ch11_12} is deterministic, so we add a stochastic term as a modeling step, following the Langevin treatment of neural noise in~\cite[Ch.~11]{ErmentroutTerman2010_ch11_12} and the use of input noise to drive spontaneous UP/DOWN switching in bistable rate models~\cite[Fig.~12.5]{ErmentroutTerman2010_ch11_12}. Here, $\xi_i(t)$ is zero-mean, unit-intensity Gaussian white noise, independent across populations, and captures fluctuations in that population's summed synaptic input. These fluctuations are dominated by the stochastic arrival of many synaptic inputs~\cite{Burkitt2006-ao}, with smaller intrinsic contributions from probabilistic neurotransmitter release, spontaneous release, and ion-channel noise~\cite{Faisal2008}. The prefactor $\sigma\sqrt{2\tau}$ makes $D=\sigma^2$ the noise intensity in the Kramers exponent of Eq.~(\ref{eq:bio_kramers}), independent of $\tau$:

\begin{equation}
    \begin{aligned}
        \tau \frac{dv_i}{dt} &= I - v_i +\sigma\sqrt{2\tau}\xi_i(t)\\
        &+ w_{\text{self}}\phi(v_i) + \wc\sum_{j \neq i}\phi(v_j),\\
        i &=1,2,3
    \end{aligned}
    \label{eq:bio_coupled}
\end{equation}

Under the assumptions below, three features of Eq.~(\ref{eq:bio_coupled}) permit a reduction to discrete dynamics. First, consider the deterministic single-population drift of an isolated unit,

\begin{equation}
    f(v_i) = I - v_i + w_\text{self}\,\phi(v_i).
    \label{eq:bio_singledrift}
\end{equation}

For a sigmoidal $\phi$ with a single inflection point, $f'(v) = -1 + w_\text{self}\,\phi'(v)$ is positive on some interval only when the self-coupling exceeds the critical value $w_c = 1/\max_v \phi'(v)$~\cite[Sec.~12.3.1]{ErmentroutTerman2010_ch11_12}. When $w_\text{self} > w_c$ and $I$ lies between the two saddle-node values at which $f$ and $f'$ vanish simultaneously, $f$ has three roots: two stable fixed points $v_-^\ast$, $v_+^\ast$, separated by a barrier $v_b^\ast$. The mean somatic potential $v_i$ of each population, therefore, resides, over long timescales, in one of two basins, and we label this choice by a binary variable $b_i \in \{0,1\}$. The effective potential $U(v)=-\!\int^v f(v')\,\,dv'$ implied by Eq.~(\ref{eq:bio_singledrift}) is the double well that governs transitions. This double well has a basis in the biological literature, with Loewenstein et al. showing that cerebellar Purkinje cells are bistable due to the cell's ion channels and Sanchez-Vives and McCormick showing that cortical slices generate UP/DOWN states~\cite{Loewenstein2005, Sanchez-Vives2000}.

Second, noise drives transitions across the barrier $v_b^\ast$ at the Kramers rate,

\begin{equation}
  \begin{aligned}
    \gamma &= \frac{1}{2\pi\tau} \sqrt{\left | U'' \left (v_-^{\ast} \right) \; U'' \left (v_{b}^{\ast} \right) \right |} \; \exp \! \left ( - \frac{\Delta U}{D} \right )\\
    D &= \sigma^2
  \end{aligned}
  \label{eq:bio_kramers}
\end{equation}

\noindent with $\Delta U = U \left (v_{b}^{\ast} \right) - U \left (v_-^{\ast} \right)$. $\gamma$ is the symbolic rate defined by Eq.~(\ref{eq:bio_kramers}), and represents the rate of spontaneous noise-driven switching of a population between its low- and high-activity states, the model analog of UP/DOWN transitions. It depends on $w_\text{self}$, $I$, and $\phi$ through the barrier height $\Delta U$ and the curvatures $U''$, and on the noise intensity $D$. These parameters are not required for the structural result, Eq.~(\ref{eq:bio_master}).

Finally, the reduction to a discrete configuration is valid in the metastable regime, where relaxation within a basin is fast compared to inter-basin transitions, $\tau_\text{relax} \ll \gamma^{-1},\,\kappa^{-1}$. Then, the continuous states of the populations collapse into discrete configurations $(b_1, b_2, b_3) \in \{0,1\}^3$, and the dynamics are a Markov jump process between these eight states. Reducing the continuous activity to a few discrete states is supported by Inagaki et al., who find that a mouse's anterior lateral motor cortex exhibits sets of discrete attractors~\cite{Inagaki2019-fe}.

Because we take the noise terms $\xi_i(t)$ in Eq.~(\ref{eq:bio_coupled}) to be independent, noise-driven, single-unit flips occur independently, and the noise-driven part of the generator factorizes into a sum of independent single-unit flip operators, $\gamma \sum_i (\pi_i - \mathbb{I})$. This independence is an approximation, consistent with the weak noise correlations measured between cortical neurons~\cite{Cohen2011, doi:10.1126/science.1179867}. Between populations, $\xi_i$ and $\xi_j$ are correlated to the extent that the populations receive shared input, which would produce simultaneous flips that this generator excludes. Within each population, the rate description itself assumes approximately independent firing; Ermentrout and Terman~\cite[Sec.~12.1.3]{ErmentroutTerman2010_ch11_12} note that even small amounts of shared input can build into synchrony across processing stages, so this assumption can also fail.

Flips are corrected by inter-unit coupling. When population $i$ is in the minority and the majority units are static, the term $\wc\sum_{j \neq i}\phi(v_j)$ in Eq.~(\ref{eq:bio_coupled}) adds a constant tilt to its potential,

\begin{equation}
  U_c = U(v) - \left ( \wc\sum_{j \neq i}\phi(v^\ast_j) \right ) v
  \label{eq:bio_tilted}
\end{equation}

\noindent lowering the barrier towards the majority basin and raising it towards the minority one. We treat the majority populations as static when evaluating the minority population's escape. Populations in the majority sit deep in wells maintained by recurrent self-excitation and, after a flip, are themselves pulled toward the shared consensus by $\wc$, so they change slowly compared with the minority population’s escape. The resulting biased escape rate back to consensus defines $\kappa$, the Kramers rate of the tilted potential in Eq.~(\ref{eq:bio_tilted}), in parallel with $\gamma$ in Eq.~(\ref{eq:bio_kramers}). In this reduced model, $\wc$ performs the role played by the recovery channel in Eq.~(\ref{eq:qec_rep_continuous_generator}): it is a directed relaxation towards the code subspace. Collecting the independent flip generator and the majority-pull generator gives

\begin{equation}
  \frac{dp}{dt} = \gamma \sum_{i=1}^3 \left ( \pi_i - \mathbb{I} \right ) p + \kappa \left ( R\ub - \mathbb{I} \right ) p
  \label{eq:bio_master}
\end{equation}

\noindent where $R\ub$ is a majority-vote recovery generator. Eq.~(\ref{eq:bio_master}) is the classical counterpart of Eq.~\eqref{eq:qec_rep_continuous_generator}: $\pi_i$ replaces $X_i \rho X_i$, and $R\ub$ replaces the recovery channel $\left ( \mathcal R\uq(\rho)-\rho \right )$, with $\gamma$ and $\kappa$ now classical rates rather than the physical bit-flip and recovery rates. The plots in Fig.~\ref{fig:bec_repetition_demo} show how these equations' results parallel their quantum counterparts in Fig.~\ref{fig:qec_repetition_demo}. To generate these plots, we define the codespace population, or the classical probability mass that sits on a valid codeword:

\begin{equation}
    P_C\ub(t) = \text{Pr}[b(t)=000] + \text{Pr}[b(t)=111]
    \label{eq:bio_codespace_pop}
\end{equation}

\noindent and the probability that constraints are violated:

\begin{equation}
    V\ub = \text{Pr}(b_1 \neq b_2) + \text{Pr}(b_2 \neq b_3)
    \label{eq:bio_constraint}
\end{equation}

Eq.~(\ref{eq:bio_codespace_pop}) is different from Eq.~(\ref{eq:qec_rep_fidelity}) due to the classical nature of neural activity, which has no quantum amplitude or coherence. Eq.~(\ref{eq:bio_constraint}) can be evaluated at the stationary distribution to get $V_\infty\ub$, similar to Eq.~(\ref{eq:qec_rep_v_infty}). By the permutation symmetry and global bit-flip symmetry of Eq.~\eqref{eq:bio_master}, the stationary distribution assigns equal weight $a$ to the two codewords and equal weight $b$ to each of the six non-codeword configurations. The stationarity condition for the consensus state $000$ balances its noise-driven outflow $3\gamma a$ against the combined noise and recovery inflow $3(\gamma+\kappa)b$ from the weight-one states, giving $\gamma a = (\gamma+\kappa)b$. With the normalization $2a+6b=1$, this fixes $b = \gamma/[2(\kappa+4\gamma)]$. Each adjacent check is violated by exactly four configurations, each carrying probability $b$, so

\begin{equation}
    \begin{aligned}
        V_\infty\ub &= \Pr[b_1 \neq b_2] + \Pr[b_2 \neq b_3] = 8b\\
        &= \frac{4\gamma}{\kappa + 4\gamma}
    \end{aligned}
    \label{eq:bio_v_infty}
\end{equation}

\noindent matching the quantum steady-state violation $V_\infty\uq$ of Eq.~(\ref{eq:qec_rep_v_infty}) for this reduced model. The two generators therefore share the same off-manifold residual under persistent noise, governed by the single ratio $\gamma/\kappa$. Li et al. optogenetically perturbed a mouse's premotor cortex, and they found that the network recovered at a finite rate~\cite{Li2016-bg}. In this test, recovery involves large populations across both hemispheres, so it is not a measure of $\kappa$ in this three-unit model.

Following Ermentrout and Terman~\cite[Sec.~12.1.3]{ErmentroutTerman2010_ch11_12}, each unit is a small, local population. The model, therefore, does not support a one-to-one mapping of single neurons onto qubits. Eq.~\eqref{eq:bio_et} is a rate model with no threshold or reset, so the results depend only on a one-dimensional double-well Langevin reduction. The same results hold for any bistable elements with majority coupling. What makes the model neural is its mechanism, recurrent excitation and synaptic noise sources, together with the experimental counterparts cited above. The units are purely excitatory, a minimal choice, and the independent noise is an approximation, as correlated input would produce correlated flips, the counterpart of correlated errors in QEC. The rates $\gamma$ and $\kappa$ remain purely symbolic, and the cited experiments are qualitative counterparts, not parameter measurements.

\section{Conclusions}

This work argues that quantum error correction and biological error correction in neuronal circuits share a common organizational pattern: redundant encoding of protected information with constraint-based inference that detects and suppresses errors. To make this analogy precise, we constructed a structural dictionary whose core objects are summarized in Table~\ref{tab:dictionary_core}: physical qubits, logical qubits, the codespace $C\uq$, noise, and code distance $d\uq$ map onto noisy physical units $x\ub_i$, low-dimensional cognitive variables $X\ub_\alpha$, a neural manifold $\mathcal{M}\ub$, neural noise, and a robustness scale $d\ub$. The extended mapping of checks, syndromes, ancillary degrees of freedom, decoding, and recovery is developed in Secs.~\ref{subsec:checks_syndromes}--\ref{subsec:decoding_correction}. The correspondence is structural rather than physical: neurons are treated throughout as noisy classical elements, and no quantum coherence or entanglement is invoked in biological tissue.

In addition to the dictionary, we show that this correspondence can be made quantitative at the level of generator structure and steady-state violation in a minimal model. Three recurrently coupled populations, described by a noisy firing-rate model~\cite[Eq.~(12.6)]{ErmentroutTerman2010_ch11_12}, reduce, in the metastable regime, to the classical master equation of Eq.~(\ref{eq:bio_master}). This generator has the same structure as the continuous-time recovery generator of the three-qubit repetition code, Eq.~(\ref{eq:qec_rep_continuous_generator}): independent bit-flip noise appears as independent single-unit flips at rate $\gamma$, and the recovery channel is replaced by a majority-pull term at rate $\kappa$ generated by the recurrent coupling $\wc$. The resulting dynamics, Fig.~\ref{fig:bec_repetition_demo}, mirror the dynamics of the quantum counterpart, Fig.~\ref{fig:qec_repetition_demo}. In particular, both systems display the same distinction between projection and damping recovery: exact correction returns the state to the codespace, whereas continuous-time recovery suppresses constraint violation, leaving a nonzero steady-state residual.

Casting both systems in this shared language suggests transfer in both directions. From biology to QEC, neural error control is local, continuous in time, and adaptive, properties that remain challenging for engineered decoders. From QEC to neuroscience, the notions of codespace, syndrome, decoder, and distance supply a quantitative vocabulary in which the representational robustness of a neural code can be defined and ultimately measured.

One caveat, however, is that the correspondence is an analogy of roles, not an identification of mechanisms: neurons are computationally far richer than qubits, and the reduction to a binary code relies on the metastable regime and on treating single-unit noise as independent. We keep the rates $\gamma$ and $\kappa$ symbolic, establishing the structural correspondence without committing to parameter identification. The quantitative model captures only the population-level neuronal dynamics, while we acknowledge that astrocytes may contribute to neuronal computation and biological error control. This is, therefore, a first step toward a quantitative correspondence rather than a final one. Natural extensions include computing the robustness scale $d^{(B)}$, identifying $\gamma$ and $\kappa$ from population recordings, incorporating correlated input and inhibition, and moving beyond three-unit codes. Future astrocytic contributions are discussed in Sec.~\ref{sec:fw}

\subsection{Future Work -- Astrocytes and Derived Algorithms}
\label{sec:fw}

In this work, we argued that astrocytes may contribute to biological error control and prevention, however, we do not use them to inform our analogy outside of their potential role as analogs of ancillary degrees of freedom. Astrocytes are the most numerous cell within the brain and may bind to between $270,000$ and 2 million synapses~\cite{freeman2013evolving, Oberheim2009}. They process information from the neurons, integrating the neurons' activations non-linearly, suggesting they may play an active computational role~\cite{PEREA2009421}. Therefore, astrocytes may play an important, but still incompletely understood role in the neuron's population coding scheme. One mechanism they may act through is by reshaping and adapting the neuronal network for specific contexts, as suggested by Murphy-Royal et al.~\cite{Murphy-Royal_Ching_Papouin_2023}.

In some ways, astrocytes may contribute to the brain's error protection. Since the brain is a subcritical system that operates close to criticality, astrocytes may help regulate the system's state by modulating transmission and plasticity~\cite{36v9-wtm8, Noriega-Prieto_Araque_2021}. Astrocytes are large and highly connected, enabling them to regulate far-away synapses, and suggesting how a network of the cells could contribute to brain stability~\cite{COVELO201662}. Finally, astrocytes have been linked to neuronal network self-repair, which suggests they may help detect faults and address them autonomously~\cite{Wade_2012}.

Astrocyte-inspired algorithms suggest a useful analogy to syndrome extraction: continuous syndrome-like monitoring. In rhythmic sharing, the oscillator phases are not themselves syndromes. Rather, deviations of synchrony or phase-locking structure from a learned baseline provide mismatch signals indicating that the current input stream is no longer compatible with the learned dynamical regime. The resulting reconfiguration is therefore closer to adaptive constraint monitoring than to projective stabilizer readout. For example, the Rhythmic Sharing algorithm modulates effective synaptic coupling via slow oscillations driven by astrocytic rhythms~\cite{fdc2-ljj6}. Such rhythms can provide otherwise static recurrent networks with adaptive performance and sensitivity to distributional shift~\cite{WhitehouseKangLosert2025}.

Another algorithm based on astrocytes, called $\vec{\delta}$-Multiplexed Gradient Descent, uses an astrocyte-inspired module to change an artificial neural network's weights, leading to a biologically-plausible alternative to backpropagation~\cite{dmgd}. This suggests how astrocyte-like variables could help establish population codes for complex neuronal problem-solving. In quantum, logical information is encoded into highly entangled subspaces of many physical qubits; ancillary qubits are used to measure stabilizer generators (parity checks); algorithms map syndrome patterns to correction operations~\cite{NielsenChuang2010,Terhal2015}. In biology, classical population codes are implemented by neurons, with astrocytes potentially supplying additional degrees of freedom that monitor, average, and modulate neural activity. We plan to carry out future work to understand how astrocytes fit within the analogy we have discussed here.

\section*{Declarations}

\bmhead{Acknowledgments}

This study was funded in part by the Air Force Office of Scientific Research Biophysics Program [Grants \textit{FA9550-22-1-0405} and \textit{FA9550-25-1-0002}]. The funder played no role in study design, data collection, analysis and interpretation of data, or the writing of this manuscript.

\bmhead{Author Contributions}

IW and AZ found initial background research and provided derivations for the neuroscience and quantum sections, respectively. FK and WL supervised and provided insight from quantum computing and wetlab experiments. MEAM, WS, SC, SD contributed to background research and the structural dictionary. All authors reviewed the manuscript.

\bmhead{Competing Interests}

The authors declare no competing financial or non-financial interests.

% Bibliography
% \bibliographystyle{apsrev4-2}
\bibliography{refs} 

@incollection{vonNeumann1956,
	title        = {Probabilistic Logics and the Synthesis of Reliable Organisms from Unreliable Components},
	author       = {von Neumann, John},
	year         = 1956,
	booktitle    = {Automata Studies},
	publisher    = {Princeton University Press},
	address      = {Princeton, NJ},
	series       = {Annals of Mathematics Studies},
	volume       = 34,
	pages        = {43--98},
	editor       = {Shannon, Claude E. and McCarthy, John}
}

@article{Shor1995,
	title        = {Scheme for reducing decoherence in quantum computer memory},
	author       = {Shor, Peter W.},
	year         = 1995,
	journal      = {Physical Review A},
	volume       = 52,
	number       = 4,
	pages        = {R2493--R2496},
	doi          = {10.1103/PhysRevA.52.R2493}
}

@book{NielsenChuang2010,
	title        = {Quantum Computation and Quantum Information},
	author       = {Nielsen, Michael A. and Chuang, Isaac L.},
	year         = 2010,
	publisher    = {Cambridge University Press},
	address      = {Cambridge},
	edition      = {10th Anniversary Edition}
}

@article{Terhal2015,
	title        = {Quantum error correction for quantum memories},
	author       = {Terhal, Barbara M.},
	year         = 2015,
	journal      = {Reviews of Modern Physics},
	volume       = 87,
	number       = 2,
	pages        = {307--346},
	doi          = {10.1103/RevModPhys.87.307}
}

@article{BurakFiete2009,
	title        = {Accurate path integration in continuous attractor network models of grid cells},
	author       = {Burak, Yoram and Fiete, Ila R.},
	year         = 2009,
	journal      = {PLoS Computational Biology},
	volume       = 5,
	number       = 2,
	pages        = {e1000291},
	doi          = {10.1371/journal.pcbi.1000291}
}

@article{ChaudhuriFiete2016,
	title        = {Computational principles of memory},
	author       = {Chaudhuri, Rishidev and Fiete, Ila R.},
	year         = 2016,
	journal      = {Nature Neuroscience},
	volume       = 19,
	pages        = {394--403},
	doi          = {10.1038/nn.4237}
}

@article{SreenivasanFiete2011,
	title        = {Grid cells generate an analog error-correcting code for singularly precise neural computation},
	author       = {Sreenivasan, Sameet and Fiete, Ila R.},
	year         = 2011,
	journal      = {Nature Neuroscience},
	volume       = 14,
	number       = 10,
	pages        = {1330--1337},
	doi          = {10.1038/nn.2901}
}

@article{Zlokapa2024,
	title        = {Fault-tolerant neural networks from biological error correction codes},
	author       = {Zlokapa, Alexander and Tan, Andrew K. and Martyn, John M. and Fiete, Ila R. and Tegmark, Max and Chuang, Isaac L.},
	year         = 2024,
	journal      = {Physical Review E},
	volume       = 110,
	number       = 5,
	pages        = {054303},
	doi          = {10.1103/PhysRevE.110.054303}
}

@article{Santello2019,
	title        = {Astrocyte function from information processing to cognition and cognitive impairment},
	author       = {Santello, Mirko and Toni, Nicolas and Volterra, Andrea},
	year         = 2019,
	journal      = {Nature Neuroscience},
	volume       = 22,
	number       = 2,
	pages        = {154--166},
	doi          = {10.1038/s41593-018-0325-8}
}

@article{KofujiAraque2021,
	title        = {Astrocytes and behavior},
	author       = {Kofuji, Paulo and Araque, Alfonso},
	year         = 2021,
	journal      = {Annual Review of Neuroscience},
	volume       = 44,
	pages        = {49--67},
	doi          = {10.1146/annurev-neuro-101920-112225}
}

@article{ONeill2023,
	title        = {Decoding natural astrocyte rhythms: Dynamic actin waves result from environmental sensing by primary rodent astrocytes},
	author       = {O'Neill, K. M. and others},
	year         = 2023,
	journal      = {Advanced Biology},
	volume       = 7,
	number       = 6,
	pages        = 2200269,
	doi          = {10.1002/adbi.202200269}
}

@article{Lee2014,
	title        = {Astrocytes contribute to gamma oscillations and recognition memory},
	author       = {Lee, H. and others},
	year         = 2014,
	journal      = {Proceedings of the National Academy of Sciences of the USA},
	volume       = 111,
	number       = 32,
	pages        = {E3343--E3352},
	doi          = {10.1073/pnas.1410893111}
}

@article{JOHNSON2000563,
	title        = {Neural Coding},
	author       = {Kenneth O Johnson},
	year         = 2000,
	journal      = {Neuron},
	volume       = 26,
	number       = 3,
	pages        = {563--566},
	doi          = {https://doi.org/10.1016/S0896-6273(00)81193-9},
	issn         = {0896-6273},
	url          = {https://www.sciencedirect.com/science/article/pii/S0896627300811939}
}

@misc{Li_2024,
	title        = {Revealing unexpected complex encoding but simple decoding mechanisms in motor cortex via separating behaviorally relevant neural signals},
	author       = {Li, Yangang and Zhu, Xinyun and Qi, Yu and Wang, Yueming},
	year         = 2024,
	month        = jul,
	publisher    = {eLife Sciences Publications, Ltd},
	doi          = {10.7554/elife.87881.3},
    journal      = {eLife},
	abstractnote = {<jats:p>Abstract</jats:p> <jats:p>In motor cortex, behaviorally-relevant neural responses are entangled with irrelevant signals, which complicates the study of encoding and decoding mechanisms. It remains unclear whether behaviorally-irrelevant signals could conceal some critical truth. One solution is to accurately separate behaviorally-relevant and irrelevant signals at both single-neuron and single-trial levels, but this approach remains elusive due to the unknown ground truth of behaviorally-relevant signals. Therefore, we propose a framework to define, extract, and validate behaviorally-relevant signals. Analyzing separated signals in three monkeys performing different reaching tasks, we found neural responses previously considered to contain little information actually encode rich behavioral information in complex nonlinear ways. These responses are critical for neuronal redundancy and reveal movement behaviors occupy a higher-dimensional neural space than previously expected. Surprisingly, when incorporating often-ignored neural dimensions, behaviorally-relevant signals can be decoded linearly with comparable performance to nonlinear decoding, suggesting linear readout may be performed in motor cortex. Our findings prompt that separating behaviorally-relevant signals may help uncover more hidden cortical mechanisms.</jats:p>}
}

@article{Safaai2025,
	title        = {Specialized structure of neural population codes in parietal cortex outputs},
	author       = {Safaai,  Houman and Wang,  Alice Y. and Kira,  Shinichiro and Blanco Malerba,  Simone and Panzeri,  Stefano and Harvey,  Christopher D.},
	year         = 2025,
	month        = oct,
	journal      = {Nature Neuroscience},
	publisher    = {Springer Science and Business Media LLC},
	volume       = 28,
	number       = 12,
	pages        = {2550–2560},
	doi          = {10.1038/s41593-025-02095-x},
	issn         = {1546-1726},
}

@article{Xie2024,
	title        = {Neuronal sequences in population bursts encode information in human cortex},
	author       = {Xie,  Weizhen and Wittig,  John H. and Chapeton,  Julio I. and El-Kalliny,  Mostafa and Jackson,  Samantha N. and Inati,  Sara K. and Zaghloul,  Kareem A.},
	year         = 2024,
	month        = oct,
	journal      = {Nature},
	publisher    = {Springer Science and Business Media LLC},
	volume       = 635,
	number       = 8040,
	pages        = {935–942},
	doi          = {10.1038/s41586-024-08075-8},
	issn         = {1476-4687},
}

@article{doi:10.1073/pnas.1004906107,
	title        = {Optimal population coding by noisy spiking neurons},
	author       = {Gašper Tkačik  and Jason S. Prentice  and Vijay Balasubramanian  and Elad Schneidman},
	year         = 2010,
	journal      = {Proceedings of the National Academy of Sciences},
	volume       = 107,
	number       = 32,
	pages        = {14419--14424},
	doi          = {10.1073/pnas.1004906107},
	url          = {https://www.pnas.org/doi/abs/10.1073/pnas.1004906107}
}

@article{Boerlin2013,
	title        = {Predictive Coding of Dynamical Variables in Balanced Spiking Networks},
	author       = {Boerlin,  Martin and Machens,  Christian K. and Denève,  Sophie},
	year         = 2013,
	month        = nov,
	journal      = {PLoS Computational Biology},
	publisher    = {Public Library of Science (PLoS)},
	volume       = 9,
	number       = 11,
	pages        = {e1003258},
	doi          = {10.1371/journal.pcbi.1003258},
	issn         = {1553-7358},
	editor       = {Sporns,  Olaf}
}

@inproceedings{NIPS2012_3a15c7d0,
	title        = {Learning optimal spike-based representations},
	author       = {Bourdoukan, Ralph and Barrett, David and Deneve, Sophie and Machens, Christian K},
	year         = 2012,
	booktitle    = {Advances in Neural Information Processing Systems},
	publisher    = {Curran Associates, Inc.},
	volume       = 25,
	pages        = {},
	url          = {https://proceedings.neurips.cc/paper_files/paper/2012/file/3a15c7d0bbe60300a39f76f8a5ba6896-Paper.pdf},
	editor       = {F. Pereira and C.J. Burges and L. Bottou and K.Q. Weinberger}
}

@article{10.3389/fncom.2020.00020,
	title        = {Clustering of Neural Activity: A Design Principle for Population Codes},
	author       = {Berry, Michael J.  and Tkačik, Gašper},
	year         = 2020,
	journal      = {Frontiers in Computational Neuroscience},
	volume       = {Volume 14 - 2020},
	doi          = {10.3389/fncom.2020.00020},
	issn         = {1662-5188},
	url          = {https://www.frontiersin.org/journals/computational-neuroscience/articles/10.3389/fncom.2020.00020}
}

@article{36v9-wtm8,
	title        = {Robust Scaling in Human Brain Dynamics Despite Correlated Inputs and Limited Sampling Distortions},
	author       = {Calvo, Rub\'en and Martorell, Carles and Roig, Adri\'an and Mu\~noz, Miguel A.},
	year         = 2026,
	month        = {Feb},
	journal      = {Phys. Rev. Lett.},
	publisher    = {American Physical Society},
	volume       = 136,
	pages        = {068402},
	doi          = {10.1103/36v9-wtm8},
	url          = {https://link.aps.org/doi/10.1103/36v9-wtm8},
	issue        = 6,
	numpages     = 8
}

@article{10.1098/rstb.2018.0377,
	title        = {Evolutionary aspects of reservoir computing},
	author       = {Seoane, Luís F.},
	year         = 2019,
	month        = {04},
	journal      = {Philosophical Transactions of the Royal Society B: Biological Sciences},
	volume       = 374,
	number       = 1774,
	pages        = 20180377,
	doi          = {10.1098/rstb.2018.0377},
	issn         = {0962-8436},
}

@article{10.7554/eLife.73276,
	title        = {The geometry of robustness in spiking neural networks},
	author       = {Calaim, Nuno and Dehmelt, Florian A and Gonçalves, Pedro J and Machens, Christian K},
	year         = 2022,
	month        = {may},
	journal      = {eLife},
	publisher    = {eLife Sciences Publications, Ltd},
	volume       = 11,
	pages        = {e73276},
	doi          = {10.7554/eLife.73276},
	issn         = {2050-084X},
	url          = {https://doi.org/10.7554/eLife.73276},
	article_type = {journal},
	editor       = {Meister, Markus and Frank, Michael J and Doiron, Brent and Meister, Markus},
	pub_date     = {2022-05-30},
	citation     = {eLife 2022;11:e73276}
}

@inbook{Arle2020,
	title        = {Robustness in Neural Circuits},
	author       = {Arle,  Jeffrey E. and Mei,  Longzhi and Carlson,  Kristen W.},
	year         = 2020,
	month        = aug,
	booktitle    = {Brain and Human Body Modeling 2020},
	publisher    = {Springer International Publishing},
	pages        = {213–229},
	doi          = {10.1007/978-3-030-45623-8_12},
	isbn         = 9783030456238,
}

@article{Lim2013,
	title        = {Balanced cortical microcircuitry for maintaining information in working memory},
	author       = {Lim,  Sukbin and Goldman,  Mark S},
	year         = 2013,
	month        = aug,
	journal      = {Nature Neuroscience},
	publisher    = {Springer Science and Business Media LLC},
	volume       = 16,
	number       = 9,
	pages        = {1306–1314},
	doi          = {10.1038/nn.3492},
	issn         = {1546-1726},
}

@article{adams2020quantum,
	title        = {Quantum effects in the brain: A review},
	author       = {Adams, Betony and Petruccione, Francesco},
	year         = 2020,
	journal      = {AVS Quantum Science},
	publisher    = {AIP Publishing},
	volume       = 2,
	number       = 2
}

@article{neven2024testing,
	title        = {Testing the conjecture that quantum processes create conscious experience},
	author       = {Neven, Hartmut and Zalcman, Adam and Read, Peter and Kosik, Kenneth S and van der Molen, Tjitse and Bouwmeester, Dirk and Bodnia, Eve and Turin, Luca and Koch, Christof},
	year         = 2024,
	journal      = {Entropy},
	publisher    = {MDPI},
	volume       = 26,
	number       = 6,
	pages        = 460
}

@article{wahbeh2022if,
	title        = {What if consciousness is not an emergent property of the brain? {Observational} and empirical challenges to materialistic models},
	author       = {Wahbeh, Helan{\'e} and Radin, Dean and Cannard, Cedric and Delorme, Arnaud},
	year         = 2022,
	journal      = {Frontiers in Psychology},
	publisher    = {Frontiers},
	volume       = 13,
	pages        = 955594
}

@article{derakhshani2022crossroad,
	title        = {At the crossroad of the search for spontaneous radiation and the {Orch OR} consciousness theory},
	author       = {Derakhshani, Maaneli and Di{\'o}si, Lajos and Laubenstein, Matthias and Piscicchia, Kristian and Curceanu, Catalina},
	year         = 2022,
	journal      = {Physics of Life Reviews},
	publisher    = {Elsevier},
	volume       = 42,
	pages        = {8--14}
}

@inproceedings{craddock2025quantum,
	title        = {Quantum mechanisms in the brain: from conjectures and theories to experimental evidence},
	author       = {Craddock, Travis JA},
	year         = 2025,
	booktitle    = {Quantum Effects and Measurement Techniques in Biology and Biophotonics II},
	volume       = 13340,
	pages        = {1334003},
	organization = {SPIE},
}

@article{Oberheim2009,
	title        = {Uniquely Hominid Features of Adult Human Astrocytes},
	author       = {Oberheim,  Nancy Ann and Takano,  Takahiro and Han,  Xiaoning and He,  Wei and Lin,  Jane H. C. and Wang,  Fushun and Xu,  Qiwu and Wyatt,  Jeffrey D. and Pilcher,  Webster and Ojemann,  Jeffrey G. and Ransom,  Bruce R. and Goldman,  Steven A. and Nedergaard,  Maiken},
	year         = 2009,
	month        = mar,
	journal      = {The Journal of Neuroscience},
	publisher    = {Society for Neuroscience},
	volume       = 29,
	number       = 10,
	pages        = {3276–3287},
	doi          = {10.1523/jneurosci.4707-08.2009},
	issn         = {1529-2401},
}

@article{PEREA2009421,
	title        = {Tripartite synapses: astrocytes process and control synaptic information},
	author       = {Gertrudis Perea and Marta Navarrete and Alfonso Araque},
	year         = 2009,
	journal      = {Trends in Neurosciences},
	volume       = 32,
	number       = 8,
	pages        = {421--431},
	doi          = {https://doi.org/10.1016/j.tins.2009.05.001},
	issn         = {0166-2236},
	url          = {https://www.sciencedirect.com/science/article/pii/S0166223609001015}
}

@article{Murphy-Royal_Ching_Papouin_2023,
	title        = {A conceptual framework for astrocyte function},
	author       = {Murphy-Royal, Ciaran and Ching, ShiNung and Papouin, Thomas},
	year         = 2023,
	month        = oct,
	journal      = {Nature Neuroscience},
	publisher    = {Springer Science and Business Media LLC},
	volume       = 26,
	number       = 11,
	pages        = {1848–1856},
	doi          = {10.1038/s41593-023-01448-8},
	language     = {en}
}

@article{Noriega-Prieto_Araque_2021,
	title        = {Sensing and Regulating Synaptic Activity by Astrocytes at Tripartite Synapse},
	author       = {Noriega-Prieto, José Antonio and Araque, Alfonso},
	year         = 2021,
	month        = apr,
	journal      = {Neurochemical Research},
	publisher    = {Springer Science and Business Media LLC},
	volume       = 46,
	number       = 10,
	pages        = {2580–2585},
	doi          = {10.1007/s11064-021-03317-x},
	language     = {en}
}

@article{COVELO201662,
	title        = {Lateral regulation of synaptic transmission by astrocytes},
	author       = {A. Covelo and A. Araque},
	year         = 2016,
	journal      = {Neuroscience},
	volume       = 323,
	pages        = {62--66},
	doi          = {https://doi.org/10.1016/j.neuroscience.2015.02.036},
	issn         = {0306-4522},
	url          = {https://www.sciencedirect.com/science/article/pii/S0306452215001888},
	note         = {Dynamic and metabolic astrocyte-neuron interactions in healthy and diseased brain}
}

@article{Wade_2012,
	title        = {Self-repair in a bidirectionally coupled astrocyte-neuron {(AN)} system based on retrograde signaling},
	author       = {Wade, John},
	year         = 2012,
	journal      = {Frontiers in Computational Neuroscience},
	publisher    = {Frontiers Media SA},
	volume       = 6,
	doi          = {10.3389/fncom.2012.00076},
}

@phdthesis{Gottesman1997,
  author = {Daniel Gottesman},
  title  = {Stabilizer Codes and Quantum Error Correction},
  school = {California Institute of Technology},
  year   = {1997},
  note   = {arXiv:quant-ph/9705052},
  doi    = {10.48550/arXiv.quant-ph/9705052}
}

@article{AhnDohertyLandahl2002,
  author  = {Charlene Ahn and Andrew C. Doherty and Andrew J. Landahl},
  title   = {Continuous Quantum Error Correction via Quantum Feedback Control},
  journal = {Physical Review A},
  volume  = {65},
  number  = {4},
  pages   = {042301},
  year    = {2002},
  doi     = {10.1103/PhysRevA.65.042301}
}

@article{SarovarMilburn2005,
  author  = {Mohan Sarovar and G. J. Milburn},
  title   = {Continuous Quantum Error Correction by Cooling},
  journal = {Physical Review A},
  volume  = {72},
  number  = {1},
  pages   = {012306},
  year    = {2005},
  doi     = {10.1103/PhysRevA.72.012306}
}

@article{KerckhoffEtAl2010,
  author  = {Joseph Kerckhoff and Hendra I. Nurdin and Dmitri S. Pavlichin and Hideo Mabuchi},
  title   = {Designing Quantum Memories with Embedded Control: Photonic Circuits for Autonomous Quantum Error Correction},
  journal = {Physical Review Letters},
  volume  = {105},
  number  = {4},
  pages   = {040502},
  year    = {2010},
  doi     = {10.1103/PhysRevLett.105.040502}
}

@article{WhitehouseKangLosert2025,
    title = {Emergent detection of concept drift within the glia-inspired ‘rhythmic sharing’ algorithm},
    volume = {3},
    ISSN = {3004-8672},
    url = {http://dx.doi.org/10.1038/s44335-026-00067-3},
    DOI = {10.1038/s44335-026-00067-3},
    number = {1},
    journal = {npj Unconventional Computing},
    publisher = {Springer Science and Business Media LLC},
    author = {Whitehouse,  Ian and Kang,  Hoony and Losert,  Wolfgang},
    year = {2026},
    month = Jun
}

@article{freeman2013evolving,
  title={Evolving concepts of gliogenesis: a look way back and ahead to the next 25 years},
  author={Freeman, Marc R and Rowitch, David H},
  journal={Neuron},
  volume={80},
  number={3},
  pages={613--623},
  year={2013},
  publisher={Elsevier}
}

@ARTICLE{Burkitt2006-ao,
  title     = "A review of the integrate-and-fire neuron model: {II}.
               Inhomogeneous synaptic input and network properties",
  author    = "Burkitt, A N",
  journal   = "Biol. Cybern.",
  publisher = "Springer Science and Business Media LLC",
  volume    =  95,
  number    =  2,
  pages     = "97--112",
  month     =  aug,
  year      =  2006,
  language  = "en"
}

@article{BENIAGUEV20212727,
title = {Single cortical neurons as deep artificial neural networks},
journal = {Neuron},
volume = {109},
number = {17},
pages = {2727-2739.e3},
year = {2021},
issn = {0896-6273},
doi = {https://doi.org/10.1016/j.neuron.2021.07.002},
url = {https://www.sciencedirect.com/science/article/pii/S0896627321005018},
author = {David Beniaguev and Idan Segev and Michael London}
}

@article {Aizenbud2026.06.08.730984,
	author = {Aizenbud, Ido and Beniaguev, David and Pnueli, Noam and Segev, Idan and London, Michael},
	title = {What can a neuron compute},
	elocation-id = {2026.06.08.730984},
	year = {2026},
	doi = {10.64898/2026.06.08.730984},
	publisher = {Cold Spring Harbor Laboratory},
	URL = {https://www.biorxiv.org/content/early/2026/06/09/2026.06.08.730984},
	journal = {bioRxiv}
}

@article{fdc2-ljj6,
  title = {Rhythmic sharing: A bioinspired paradigm for zero-shot adaptive learning in neural networks},
  author = {Kang, Hoony and Losert, Wolfgang},
  journal = {Phys. Rev. Res.},
  volume = {8},
  issue = {1},
  pages = {013267},
  numpages = {11},
  year = {2026},
  month = {Mar},
  publisher = {American Physical Society},
  doi = {10.1103/fdc2-ljj6},
  url = {https://link.aps.org/doi/10.1103/fdc2-ljj6}
}

@INPROCEEDINGS{dmgd,
  author={O'Loughlin, Ryan and Oripov, Bakhrom and Skuda, Nick and Chongsiriwatana, Noah and Whitehouse, Ian and Losert, Wolfgang and Hayes, Bradley and McCaughan, Adam and Buckley, Sonia},
  booktitle={2026 Neuro Inspired Computational Elements (NICE)}, 
  title={$\vec{\delta}$ Multiplexed Gradient Descent: Perturbative Learning with Astrocytes}, 
  year={2026},
  volume={},
  number={},
  pages={1--9},
  doi={10.1109/NICE69539.2026.11567454}}

@ARTICLE{Barranca2026-xp,
  title     = "Distance-dependent connectivity in the brain facilitates high
               dynamical and structural complexity",
  author    = "Barranca, Victor J",
  journal   = "Cogn. Neurodyn.",
  publisher = "Springer Science and Business Media LLC",
  volume    =  20,
  number    =  1,
  pages     = "23",
  month     =  dec,
  year      =  2026,
  copyright = "https://creativecommons.org/licenses/by/4.0",
  language  = "en"
}

@book{hairer,
  title = {Solving Ordinary Differential Equations II},
  ISBN = {9783662099476},
  ISSN = {0179-3632},
  url = {http://dx.doi.org/10.1007/978-3-662-09947-6},
  DOI = {10.1007/978-3-662-09947-6},
  journal = {Springer Series in Computational Mathematics},
  publisher = {Springer Berlin Heidelberg},
  author = {Hairer,  Ernst and Wanner,  Gerhard},
  year = {1991}
}

@Article{Loewenstein2005,
author={Loewenstein, Yonatan
and Mahon, S{\'e}verine
and Chadderton, Paul
and Kitamura, Kazuo
and Sompolinsky, Haim
and Yarom, Yosef
and H{\"a}usser, Michael},
title={Bistability of cerebellar {Purkinje} cells modulated by sensory stimulation},
journal={Nature Neuroscience},
year={2005},
month={Feb},
day={01},
volume={8},
number={2},
pages={202-211},
issn={1546-1726},
doi={10.1038/nn1393},
url={https://doi.org/10.1038/nn1393}
}

@Article{Sanchez-Vives2000,
author={Sanchez-Vives, Maria V. and McCormick, David A.},
title={Cellular and network mechanisms of rhythmic recurrent activity in neocortex},
journal={Nature Neuroscience},
year={2000},
month={Oct},
day={01},
volume={3},
number={10},
pages={1027-1034},
issn={1546-1726},
doi={10.1038/79848},
url={https://doi.org/10.1038/79848}
}

@Article{Faisal2008,
author={Faisal, A. Aldo
and Selen, Luc P. J.
and Wolpert, Daniel M.},
title={Noise in the nervous system},
journal={Nature Reviews Neuroscience},
year={2008},
month={Apr},
day={01},
volume={9},
number={4},
pages={292-303},
issn={1471-0048},
doi={10.1038/nrn2258},
url={https://doi.org/10.1038/nrn2258}
}

@Article{Cohen2011,
author={Cohen, Marlene R.
and Kohn, Adam},
title={Measuring and interpreting neuronal correlations},
journal={Nature Neuroscience},
year={2011},
month={Jul},
day={01},
volume={14},
number={7},
pages={811-819},
issn={1546-1726},
doi={10.1038/nn.2842},
url={https://doi.org/10.1038/nn.2842}
}

@article{
doi:10.1126/science.1179867,
author = {Alexander S. Ecker  and Philipp Berens  and Georgios A. Keliris  and Matthias Bethge  and Nikos K. Logothetis  and Andreas S. Tolias },
title = {Decorrelated Neuronal Firing in Cortical Microcircuits},
journal = {Science},
volume = {327},
number = {5965},
pages = {584-587},
year = {2010},
doi = {10.1126/science.1179867},
eprint = {https://www.science.org/doi/pdf/10.1126/science.1179867}}

@article{
doi:10.1126/science.1071795,
author = {Kazu Nakazawa  and Michael C. Quirk  and Raymond A. Chitwood  and Masahiko Watanabe  and Mark F. Yeckel  and Linus D. Sun  and Akira Kato  and Candice A. Carr  and Daniel Johnston  and Matthew A. Wilson  and Susumu Tonegawa },
title = {Requirement for Hippocampal {CA3 NMDA} Receptors in Associative Memory Recall},
journal = {Science},
volume = {297},
number = {5579},
pages = {211-218},
year = {2002},
doi = {10.1126/science.1071795},
URL = {https://www.science.org/doi/abs/10.1126/science.1071795},
eprint = {https://www.science.org/doi/pdf/10.1126/science.1071795}}

@article{
doi:10.1126/science.aal4835,
author = {Sung Soo Kim  and Hervé Rouault  and Shaul Druckmann  and Vivek Jayaraman },
title = {Ring attractor dynamics in the \textit{Drosophila} central brain},
journal = {Science},
volume = {356},
number = {6340},
pages = {849-853},
year = {2017},
doi = {10.1126/science.aal4835},
URL = {https://www.science.org/doi/abs/10.1126/science.aal4835},
eprint = {https://www.science.org/doi/pdf/10.1126/science.aal4835}}

@ARTICLE{Li2016-bg,
  title     = "Robust neuronal dynamics in premotor cortex during motor
               planning",
  author    = "Li, Nuo and Daie, Kayvon and Svoboda, Karel and Druckmann, Shaul",
  journal   = "Nature",
  publisher = "Springer Science and Business Media LLC",
  volume    =  532,
  number    =  7600,
  pages     = "459--464",
  month     =  apr,
  year      =  2016,
  language  = "en"
}

@ARTICLE{Inagaki2019-fe,
  title     = "Discrete attractor dynamics underlies persistent activity in the
               frontal cortex",
  author    = "Inagaki, Hidehiko K and Fontolan, Lorenzo and Romani, Sandro and
               Svoboda, Karel",
  journal   = "Nature",
  publisher = "Springer Science and Business Media LLC",
  volume    =  566,
  number    =  7743,
  pages     = "212--217",
  month     =  feb,
  year      =  2019,
  language  = "en"
}

@article{WILSON19721,
title = {Excitatory and Inhibitory Interactions in Localized Populations of Model Neurons},
journal = {Biophysical Journal},
volume = {12},
number = {1},
pages = {1-24},
year = {1972},
issn = {0006-3495},
doi = {https://doi.org/10.1016/S0006-3495(72)86068-5},
url = {https://www.sciencedirect.com/science/article/pii/S0006349572860685},
author = {Hugh R. Wilson and Jack D. Cowan}
}

@inbook{ErmentroutTerman2010_ch11_12,
  author    = {G. Bard Ermentrout and David H. Terman},
  title     = {Mathematical Foundations of Neuroscience},
  chapter   = {11-12},
  pages     = {283--358},
  publisher = {Springer Science \& Business Media},
  year      = {2010},
  address   = {New York, NY},
  edition   = {1st},
  isbn      = {978-0-387-87708-2},
  doi       = {10.1007/978-0-387-87708-2}
}

\end{document}